# Identifying the effect of parenthood on labor force participation: A gender comparison

*By* **Seyyed Ali Z.N. MOOSAVIAN** †

**Abstract.** Identifying the factors that influence labor force participation could elucidate how individuals arrive at their labor supply decisions, whose understanding is, in turn, of crucial importance in analyzing how the supply side of the labor market functions. This paper investigates the effect of parenthood status on Labor Force Participation (LFP) decisions using an individual-level fixed-effects identification strategy. The differences across individuals and over time in having or not having children as well as being or not being in the labor force provide the variation needed to assess the association between individuals' LFP behavior and parenthood. Parenthood could have different impacts on mothers than it would on fathers. In order to look at the causal effect of maternity and paternity on LFP separately, the data is disaggregated by gender. To this end, the effect of a change in the parenthood status can be measured using individual-level fixed-effects to account for time-invariant characteristics of individuals becoming a parent. The primary data source used is the National Longitudinal Surveys (NLS). Considering the nature of LFP variable, this paper employs Binary Response Models (BRMs) to estimate LFP equations using individual-level micro data. The findings of the study show that parenthood has a negative overall effect on LFP. However, paternity has a significant positive effect on the likelihood of being in the labor force, whilst maternity has a significant negative impact of LFP. In addition, the results imply that the effect of parenthood on LFP has been fading away over time, regardless of the gender of parents. These two pieces of evidence precisely map onto the theoretical predictions made by the related mainstream economic theories (the traditional neoclassical theory of labor supply as well as Becker's household production model). These results are robust across different models specified and various estimation methods employed. These findings can contribute to the existing knowledge about the effect of parenthood on LFP decisions made in the US at an individual and behavioral level, and also aid in the shaping of economic policies and interventions to enhance the status of labor force participation in the economy. In the end, some potential threats to the identification of this causal effect, such as endogeneity of fertility, and some possible strategies to deal with those threats are discussed.

† North Carolina State University, Department of Economics, 4102 Nelson Hall, Raleigh, NC 27695, USA.
☎ . 919.515.3274  ✉ . szeytoo@ncsu.edu

# 1. Introduction

Developing a clear understanding on how individuals make choices among consumption, leisure, and household production is of crucial importance in analyzing how the supply side of the labor market functions. To this end, identifying factors that influence labor participation would shed light on how individuals make their labor supply decisions. Labor supply decisions can be roughly divided into two types: (1) whether to work at all or not, and (2) how many hours to work, conditional on working positive hours. The present paper is essentially concerned with the former type of decisions. Accordingly, each individual must make a decision as to whether or not to work. Making such a decision can be influenced by a wide variety of demographic and socioeconomic factors, which could include labor demand, potential labor market wage, individual preferences, age, education, marital status, and parenthood status.

In this paper, I will attempt to investigate the effect of parenthood status on LFP decisions using an individual-level fixed-effects identification strategy.[2] The differences across individuals and over time in having or not having children as well as being or not being in the labor force provide the variation needed to assess the association between LFP behavior and parenthood status. In the first analytical part of the paper, a difference-in-differences (DiD) analysis will be conducted to identify the effect of having a first child - which essentially turns the status of "by-choice childlessness" into the status of "parenthood" - on labor force participation. If having a first child (henceforth, defined as parenthood for the DiD analysis part of the present paper) discourages/encourages LFP, we should see a decrease/increase in the LFP rate of individuals having their first child. At first glance, it seems that parenthood could have quite different impacts on mothers than it would on fathers, in part because naturally and historically mothers are mostly responsible to bear and rear children, while fathers in such situations are responsible to work further to be able to better financially support a larger number of dependent family members. As a result of these possible differential effects, the data will be disaggregated in terms of gender so as to be able to look at the causal effect of maternity and paternity, separately, on LFP.

---

[2]. It is important to clarify at this point that the change in parenthood status (which is the key independent variable in this study) is not theoretically an exogenous source of variation. In fact, this variable is theoretically considered to be systematically associated with the true error term of the LFP equation, and as a consequence, it is regarded as an endogenous source of variation for this model. However, since there was no information on an appropriate IV variable in the data set to be used, as far as the author is aware, and because using the available variables as IV variables was not as convincing as the use of parenthood status itself, in this study, parenthood status is still employed as the key independent variable, with having an eye on the truth that there might be some threats to the identification, which will be discussed in greater detail next. Thus, the potential endogeneity of parenthood is not accounted for in this study, and instead the exogeneity of parenthood is assumed in this paper.

The fundamental idea here is to compare the responses of two groups of individuals in two different time periods to the changes occurred to their parental status, if any, between the two time periods. The first group, which is indeed the comparison group, consists of the individuals who did not have any child in either of the periods of the study. The second group, which is in fact the study group, consists of the individuals who did not have any child in the first period, but did have a child in the second period of the study. Within the framework described above, the effect of a change in the parental status can be assessed through the use of a difference-in-differences estimator. The basic idea is to compare the response of a group of individuals affected by a change in their parental status with those of another group of individuals having similar characteristics, e.g. similar age range, being married, having a college degree, and so on, but they are unaffected by the change in the parenthood status, i.e. not having children in either of the periods being studied. The second group is indeed the "comparison group." Changes in the parenthood status, in the sense of being or not being a parent, provide a chance to apply the DiD methodology to the study of labor supply behavior in a well-defined sub-population.

Whilst straightforward and informative, this simple DiD analysis does not take other factors into account. In particular, other variables – such as age, the number of years of schooling, sex, and marital status – can also have effects on the LFP decision. To deal with these factors, a multivariate regression analysis will be carried out. Furthermore, to deal with individual-specific and time-specific sources of variations, fixed-effects models along with time dummy variables included in them will be estimated. Concerning the estimation methods to be utilized, this paper employs Binary Response Models (BRMs), including the Linear Probability Model (LPM) as well as two non-linear BRMs, i.e. Logit and Probit models, to estimate LFP equations using individual-level (micro) data. Findings of this study are expected to contribute to the existing knowledge about the effect of parenthood on LFP decisions made in the US at an individual and behavioral level, and also aid in the shaping of economic policies and interventions to enhance the status of labor force participation in the economy.

The primary data source used is the National Longitudinal Surveys (NLS), which is a set of surveys designed to gather information at multiple points in time on the labor market activities and other significant life events of several groups of men and women. Another source of data that is occasionally referred to, only to see some figures and facts about the US labor market, is the US Current Population Survey (CPS), which is a joint effort between the Bureau of Labor Statistics and the Census Bureau.

The paper uses descriptive statistical analysis, inferential statistical analysis through several univariate and multivariate regressions, as well as a difference-in-difference analysis to identify the causal effect of parenthood on LFP. The focus of the present paper is limited to the impact of parenthood on LFP, assuming the exogeneity of parenthood. This paper does not account for potential endogeneity of parenthood. However, the paper raises this concern and reviews how researchers in the field have tried to reconcile this potential issue. As mentioned previously, this study might be faced with some threats to identification, such as endogeneity of fertility, selection bias, omitted variable bias, reverse causality, measurement error, and simultaneity, which may bias the findings and complicate the interpretation of the results. In the end, these potential threats to identification and some possible strategies to deal with them, especially those appropriate to address the issue of endogeneity of fertility, will be discussed.

As Cristia (2008) states, understanding how individuals optimize their labor supply decisions in response to the arrival of a child and estimating the effect of parenthood on LFP is of importance for four justifications: (1) helping us understand the increase in female labor supply since the World War II, which can be explained by delayed childbearing and reduced fertility (Goldin, 1990), (2) gaining a better understanding of the key determinants of the female-male wage gap (Goldin & Polachek, 1987; Gronau, 1988; Fuchs, 1989; Korenman & Neumark, 1992) by attributing it to the interruptions of women's work due to childbearing, (3) providing information about time inputs invested in the child (Stafford, 1987; Blau & Grossberg, 1992) by knowing the effect of childbearing on female labor supply (if declines in labor supply after childbearing correspond to increases in child care time), and finally (4) satisfying economists' scientific, intellectual curiosity to know the quantitative importance of different determinants of female LFP.

The remainder of the paper is organized as follows. In the next section, a very brief theoretical background is offered. In section three, the existing empirical literature associated with the role of parental status on LFP is reviewed. Next, in section four, the data to be used is introduced, summary statistics are reported, and the results of the DiD analysis as well as regression analyses are provided. The main discussion will follow to interpret the findings and explain the results of the study. Section five brings up potential threats to identification and some possible strategies to deal with them. Naturally, a conclusion will follow, bringing the main points together, and discussing plans for future research. Lastly, the paper will end with appendices to elaborate further on the regression models estimated.

## 2. Theoretical background

This section provides a brief review of theoretical literature on fertility, childbearing, parenthood, and labor force participation. More specifically, the economic theory of labor supply is reviewed to build up an understanding as to how childbearing impacts individuals' decisions to participate in the labor force. To this end, two related economic theories will be described in brief: (1) the traditional neoclassical theory of labor supply, and (2) Gary Becker's theory of household production. I consider specifically how each theory facilitates an understanding of the effects of children on female and male labor force participation.

Although the primary focus of the theories of labor supply is on the decision about "how many" hours of labor to supply to the labor market, but, as Casale (2003) points out, they can also be helpful for disclosing how individuals decide whether or not to participate in the labor force and also for realizing what determinants impact their labor participation decisions.³

A labor force participation decision is ultimately about how to allocate time. In the traditional neoclassical model of labor supply, this decision involves the optimal allocation of time across two broad categories of activities: work and leisure. Nested within leisure could be childcare which can be assumed to provide direct utility to the individual. Time spent working provides an indirect source of utility through the earned wage and the increased consumption, which ultimately provides utility. In this theory, the individual is the decision-making unit who will aim to maximize utility derived from both leisure and work, subject to a scare resource of time. The decision to work and how many hours to work is therefore presented as a constrained utility maximization problem (Van der Stoep, 2008; Cahuc & Zylberberg, 2004; Reynolds *et al*, 1998; Ehrenberg & Smith, 1994). Ceteris paribus, neoclassical labor supply theory predicts that women with children will have a lower probability of participating in the labor market than those with no children.

The economic theory of household production (put forth by Gary Becker) suggests that individuals allocate time across three basic activities, which include market work, household production and pure leisure. In this model, the family rather than the individual is the only agent that maximizes its own total utility (Van der Stoep, 2008; Cahuc & Zylberberg, 2004). According to this theory, the family optimally allocates each family member's time to the three activities mentioned above to maximize its total utility function subject to a family wealth constraint. Decisions about which members and for how many hours each member engages in market work or household work will ultimately depend upon the relative productivities of family members in market work or household production (Van der Stoep, 2008; Ehrenberg & Smith, 1994; Becker, 1965). As this theory suggests, women usually have a comparative advantage in childbearing, but have also developed a comparative advantage in other aspects of household production as a result of socialization, preferences, or labor market disadvantages. Furthermore, men have historically faced higher earnings opportunities, perhaps as a result of discrimination against women in the market place (Van der Stoep, 2008; Reynolds, 1998), or motherhood penalty (Miller, 2008), or the like. Therefore, Backer's model of household production predicts a sexual division of labor among family members, and that women with children will have a lower probability of participating in the labor market than those with no children.⁴

---

³ Despite this, the notion of reservation wage can help us use these economic theories in clarifying how individuals make their LFP decisions in response to their parental status. Gronau (1973) and Ehrenberg & Smith (1994) provide good discussion on this subject matter.

⁴ According to a comparison made by Van der Stoep (2008), Becker's model of household production provides a better theoretical framework to incorporate fertility behavior and

In the economic model of household production, the likelihood of participating in the labor market for a woman will decrease with providing more child services, holding everything else constant. As Van der Stoep (2008) elaborates, the price of a woman's time in non-market activities will be also dependent upon the age composition of her children (Gronau, 1973). The younger children are, the higher reservation wage their mothers will have above the prevailing market wage, which in turn lowers their probability of participating in the labor force (Gronau, 1973). However, with the passage of the time, women gradually become more experienced in providing child services as children become older. As children grow older it will also be easier to substitute paid childcare for unpaid childcare. Moreover, children themselves may also start to take part in home production as they obtain more years of schooling, decreasing the demand for a mother's time in non-market activities.[5]

To conclude the section of theoretical framework, it should be noted that both the traditional neoclassical theory of labor supply as well as Becker's household production model predict that motherhood will be negatively related to women's LFP, since it increases a woman's reservation wage. These economic theories also predict that this negative relationship should be stronger among mothers with infants and very young children.

## 3. Empirical literature review

In this section, the existing empirical literature on the effect of parenthood on LFP decisions is reviewed in brief. To do so, a select set of studies will be reviewed from the literature of labor economics. The information and knowledge gained from this part will be used to identify the existing gaps in the literature, and also to build up an understanding of how to better design the research study in the next section.

According to Cramer (1979) and Browning (1992), the relationship between parenthood and LFP has long been of interest in the social sciences. Labor economists have typically treated parenthood as a key determinant of whether women participate in the labor force, and, if so, how many hours of work they decide to supply. This is in contrast with the way that demographers think of this relationship, as they have treated women's LFP as a key determinant of their childbearing behavior.[6] In general, a well-

---

parenthood status into LFP decisions than the traditional neoclassical theory of labor supply (Van der Stoep, 2008; Willis 1973). Unlike the traditional neoclassical model of labor supply, in the model of household production, children are entered in a household utility function as a separate variable, as opposed to being nested within a broadly-defined 'leisure' variable. As a result, the trade-offs between childcare and labor supply can be modeled more intuitively.

[5] To see further discussion on this topic, you can see Gronau (1973), Becker (1985), Connelly (1992), Browning (1992), and Van der Stoep (2008).

[6] As a consequence of the recognition of this distinction, in the past three decades, a strand of literature has formed which appreciates the causal interdependency of parenthood and LFP decisions. In this empirical literature, in particular, much attention has been paid to

established literature finds a negative relationship between childbearing and LFP, especially among women with infants and very young children. However, the interpretation of this relationship is complicated by the endogeneity of childbearing behavior. Based on whether or not a study accounts for this endogeneity, studies in this arena can be distinguished into two types: (1) studies that assume exogeneity of the variable parenthood, and (2) studies that account for the endogeneity of parenthood. Accordingly, some studies assume that motherhood is exogenous and some employ family background proxies as instrumental variables (IVs). After all, as Van der Stoep (2008) reports, many studies show that the negative relationship is robust to whether motherhood is treated as exogenous or endogenous with respect to LFP. This matter will be discussed in greater detail in what follows in this section as well as in section five.

Boushey (2008), for instance, estimates the effect of motherhood on the probability of employment among women in the US in the time period of 1979 to 2005. She controls for many variables such as race, age, marital status, and the number of non-working adults in the household; however, she does not control for possible endogeneity of childbearing. She identifies a negative partial effect of having at least one child under the age of eighteen, but she showed that this effect would become more sizable if a woman had a child under the age of six. She also finds that the motherhood effect has fallen over time, which is a result consistent with research in other OECD countries (Del Boca & Locatelli, 2006). This trend can be attributed to changes in the economic constraints that women in their LFP and fertility decisions are faced with.

Cristia (2008) estimates the causal effect of a first child on female labor supply by accounting for the endogeneity of the fertility decision by focusing on a sample of women from the National Survey of Family Growth (NSFG) who sought help to get pregnant. His results show that having a first child younger than one year old reduces female employment by 26.3 percentage points.

Eckstein & Lifshitz (2011) have analyzed the trends of female employment and participation rates by estimating a traditional female dynamic labor supply model to compare the various explanations in the literature for the observed trends. Their main finding is that the rise in education levels accounts for about one-third of the increase in female employment while about 40 percent remains unexplained by observed household characteristics. They show that this unexplained portion can be empirically attributed to changes in preferences or the costs of childrearing and household maintenance.

He & Zhu (2015) investigate fertility and female LFP in the urban areas of China. To deal with the endogeneity of fertility, they exploit twin births as the instrument for the number of children. Their results suggest that whilst the OLS estimates show that having an additional child decreased female identifying this causal mechanism so as to estimate an exogenous effect of parenthood on LFP.

LFP by 6.7% and 8.5% in 1990 and 2000, respectively, their IV estimates suggest smaller and insignificant effects for the mentioned years.

By reviewing the empirical literature on the subject matter at hand, we can readily understand the fact that no single approach has been prefect and ideal. As a result, it is important for economists to accumulate empirical evidence by using diverse empirical strategies, and the present study is meant to be such an attempt to contribute to the related literature.

Despite a growing literature on female LFP in the field, almost no attention has been paid to the effect of fatherhood on LFP. Therefore, a substantial gap that still remains in the literature is exploring the differential effects of parenthood on LFP for individuals of different genders. As far as the author is concerned, no study has used the year-by-year DiD method that is to be used in this paper, which traces the LFP behaviors of the same individuals at several select points in time over the time period of the study. Thereby, the present paper provides a time profile of historical changes in the trend of the effect of parenthood on LFP, which has not been provided so comprehensively ever in the prior literature. Additionally, this paper employs various and numerous estimation methods, and this variety and plurality of estimation methods and models have been unprecedented in the prior empirical literature. These are indeed the gaps that the present paper is to fill.

## 4. Data, regressions, results, and main discussion

The data to be used in this paper are drawn from the National Longitudinal Surveys (NLSs). NLSs are a set of surveys designed to collect information on several groups of men and women at multiple points in time on the labor market activities and other significant life events. For more than four decades, NLS data have served as an essential tool for economists, sociologists, and other researchers (Bureau of Labor Statistics, 2016). One of the most widely-used samples of NLSs in the area of labor economics has been the National Longitudinal Surveys of Youth started in 1979 (henceforth, NLSY79), which is a nationally representative sample of 12,686 young men and women, who were 14-22 years old when they were first surveyed in 1979 (www.nlsinfo.org, 2016). These individuals were interviewed annually through 1994, and thereafter, they have been being interviewed on a bi-annual basis.[7]

The data on the variables of interest in the present study were gathered and created using the NLS Investigator. The variables of interest include labor force participation[8] (henceforth, LFP) status (the depended variable), parental status (the key independent variable), and other control variables such as current marital status, year of birth, sex, educational attainment, along with some additional variables such as Aptitude, Achievement &

---

[7] To gain further information on the characteristics of this dataset, you can review the following webpages: [Rerieved from]; [Retrieved from].

[8] Labor force participation includes both individuals working and individuals looking for work (actively seeking for work).

Intelligence Scores (AFQT scores), race/ethnicity, spouse labor supply, and number of children, just in case they are needed for answering further questions arisen throughout the research. Table 1 presents the basic characteristics of the sample and the summary statistics of the main variables used in this study.

At first sight, it is somewhat surprising that the pooled LFP rate for this sample of individuals has been so high, i.e. roughly 82%, which is considerably larger than this rate for the whole US economy. As figure 1 demonstrates the LFP rate for the whole US economy during the time horizon of the present study has fluctuated between 63% and 68%.

**Table 1.** *Data description*

| Variable | Definition | Mean | Std. Dev. |
|---|---|---|---|
| LFP | A dummy variable for labor force participation which equals 1 if the individual is in the labor force during that year (The dependent variable) | 0.817 | 0.387 |
| Parenthood | A dummy variable for being a parent which is equal to 1 if the individual has at least one child during that year (The key independent variable) | 0.454 | 0.498 |
| Age | The individual's age | 31.126 | 9.581 |
| Sex | A dummy variable for gender which is equal to 1 if the individual is male, and 0 if female | 0.493 | 0.500 |
| HGC | Highest grade completed as of that year, which indicates the number of years of schooling | 12.595 | 2.364 |
| Married | A dummy variable for the individual's marital status during that year | 0.428 | 0.495 |
| Hrswk_pcy | The number of hours worked during that year | 1438.310 | 1039.852 |
| Numch | The number of children | 0.861 | 1.147 |

**Note:** There were 223,557 person-year, workable observations of 12,686 young men and women over the course of a 33-year period, after having the data cleaning process, in which the observations with incomplete data were dropped from the dataset. These individuals were 14-22 years old when they were first surveyed in 1979.

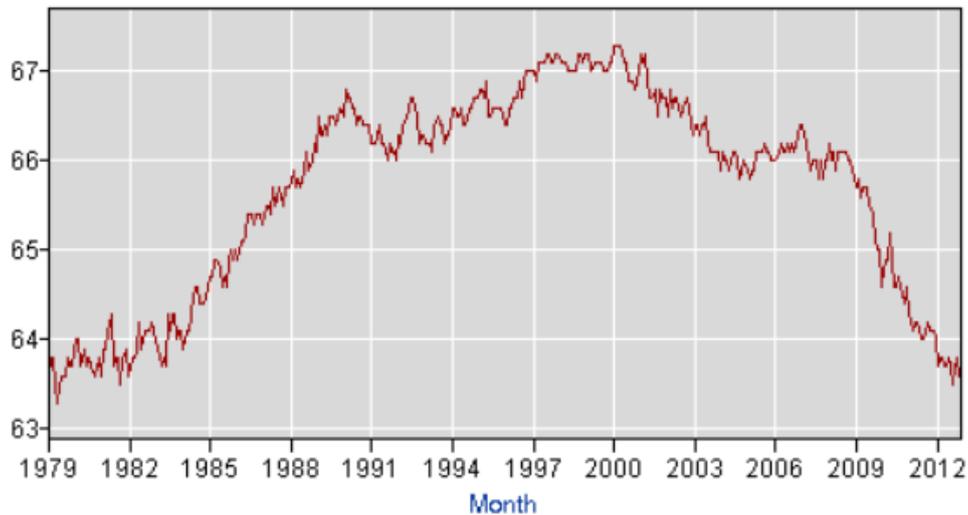

**Figure 1.** *Labor force participation rate in the US during the study period*
Extracted from the Current Population Survey at [Retrieved from].

However, further investigation on the matter clarifies that the relatively high pooled LFP rate of roughly 82% for the sample of this study should not be surprising. This is because, throughout the study period, the population under study has always been in the range of the ages at which individuals are usually very likely to be in the labor force, and as they approach the years in which they are very likely to retire, the time period of study is truncated. In other words, the truth of the matter is that there is nobody older than 55 years old in this sample. Table 2 provides a snapshot of the age structure of LFP in three separate years for the US economy. This truncation and the age structure of LFP naturally cause the pooled LFP rate of the sample of the study to be arguably somewhat larger than that of the whole US economy throughout the study horizon.[9]

**Table 2.** *Labor force participation of the US in the years of 1994, 2004, and 2014*

| Group | Participation rate | | |
|---|---|---|---|
| | 1994 | 2004 | 2014 |
| Total, 16 years and older | 66.6 | 66 | 62.9 |
| 16 to 24 | 66.4 | 61.1 | 55 |
| 20 to 24 | 77 | 75 | 70.8 |
| 25 to 54 | 83.4 | 82.8 | 80.9 |
| 25 to 34 | 83.2 | 82.7 | 81.2 |
| 35 to 44 | 84.8 | 83.6 | 82.2 |
| 45 to 54 | 81.7 | 81.8 | 79.6 |
| 55 and older | 30.1 | 36.2 | 40 |

Table 3 presents the summary statistics of the variables separately for each gender.

---

[9] To see a more complete version of table 2, you can see appendix 1.

Journal of Economics and Political Economy**Table 3.** *Means and standard deviations of variables by gender*

|  | Male | | Female | |
|---|---|---|---|---|
| # of observations | 110246 | | 113311 | |
| Variables | Mean | Standard Deviation | Mean | Standard Deviation |
| LFP | 0.859 | 0.348 | 0.775 | 0.417 |
| Parenthood | 0.344 | 0.475 | 0.560 | 0.496 |
| Age | 30.994 | 9.541 | 31.225 | 9.619 |
| HGC | 12.489 | 2.376 | 12.698 | 2.348 |
| Married | 0.400 | 0.490 | 0.456 | 0.498 |
| Hrswk_pcy | 1644.389 | 1055.884 | 1237.805 | 983.408 |
| Numch | 0.655 | 1.061 | 1.062 | 1.191 |

Table 4 presents the summary statistics of the variables separately for each parental status.

**Table 4.** *Means and standard deviations of variables by parenthood status*

|  | Non-Parents | | Parents | |
|---|---|---|---|---|
| # of observations | 122165 | | 101392 | |
| Variables | Mean | Standard Deviation | Mean | Standard Deviation |
| LFP | 0.834 | 0.372 | 0.795 | 0.404 |
| Age | 28.640 | 9.581 | 34.12 | 8.682 |
| Sex | 1.4080 | 0.491 | 1.626 | 0.484 |
| HGC | 12.603 | 2.297 | 12.584 | 2.443 |
| Married | 0.204 | 0.403 | 0.698 | 0.459 |
| Hrswk_pcy | 1394.675 | 1009.725 | 1490.885 | 1072.681 |
| Numch | 0 | 0 | 1.900 | 0.964 |

Table 5 somehow integrates the two preceding tables together and presents the summary statistics of the variables separately for each parental status and disaggregated for each gender. This matrix profile enables us to easily compare the LFP rates in terms of both gender and the status of parenthood at the same time.

**Table 5.** *A Matrix Profile Comparing the LFP Rates by Both Gender and the Status of Parenthood (Parent vs. Non-Parents)*

| LFP Rate | Male | Female |
|---|---|---|
| Parents | 0.909 | 0.727 |
|  | (0.287) | (0.445) |
|  | [37923] | [63469] |
| Non-Parents | 0.833 | 0.836 |
|  | (0.373) | (0.370) |
|  | [72323] | [49842] |

**Note:** The values bolded, values in parentheses, and values in brackets, are the LFP rates, the standard deviations, and the numbers of observations, respectively.

The above table indicates that if an individual is not a parent, then there is no significant difference between the LFP rate (or more accurately, the chance of being in the labor force) of a man and that of a women (i.e. 0.833 vs. 0.836, respectively). However, surprisingly, when the comparison group consists of individuals who are parents, then there is a huge difference between the LFP rates of different genders (i.e. 0.909 for men vs. 0.727 for women). This significant difference can be justified by having a quick look at the related theoretical literature. As discussed in section 2, the punch line of both the traditional neoclassical theory of labor supply as well as Becker's household production model is that motherhood will be negatively related to women's LFP, primarily because it increases a woman's reservation wage. In sum, we easily see that the results from the descriptive statistical analysis of the present paper clearly support the predictions of the two related economic theories.

Figure 2.a depicts the kernel densities of the number of hours worked for employed individuals before becoming a parent and after becoming a parent, conditional on being employed both before and after becoming a parent. The two kernel densities show that employed individuals, on average, work fewer hours after becoming a parent. However, figures 2.b and 2.c clarifies matters further by demonstrating the mentioned kernel densities for the case where they are disaggregated by different genders. Although the two distributions look rather similar in both panels 2.b (on the left) and 2.c (on the right), but it is absolutely obvious that the post-parenthood distribution lies considerably to the left of the pre-parenthood distribution for women, while the converse is true for the case of men, where the post-parenthood distribution lies slightly to the right of the pre-parenthood distribution. This indicates that women tend to work considerably fewer hours after becoming a parent, while men tend to work slightly more hours in this situation. Further investigations on the kernel densities show that a typical man, on average, tends to increase his hours worked by 3.5% when he becomes a parent, compared to his own hours worked when he was not a parent yet. On the other hand, it turns out that a typical woman, on average, tends to decrease her hours worked by 13.5% when she becomes a parent, compared to her own hours worked when she was not a parent yet. These patterns are all consistent with the predictions of the two mainstream related economic theories in that women tend to decrease their hours worked as they become a parent, while the converse is true for men.

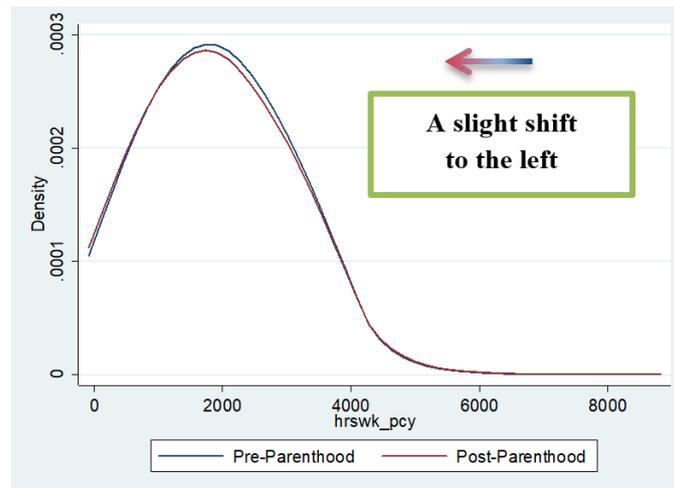

**Figure 2.a.** *Kernel Densities of the Number of Hours Worked for Employed Individuals before Becoming Parents and after Becoming Parents, Conditional on Being Employed before and after Becoming Parents*

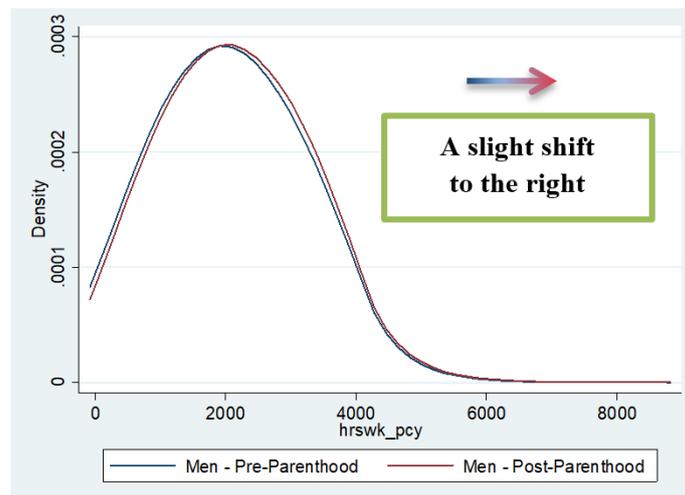

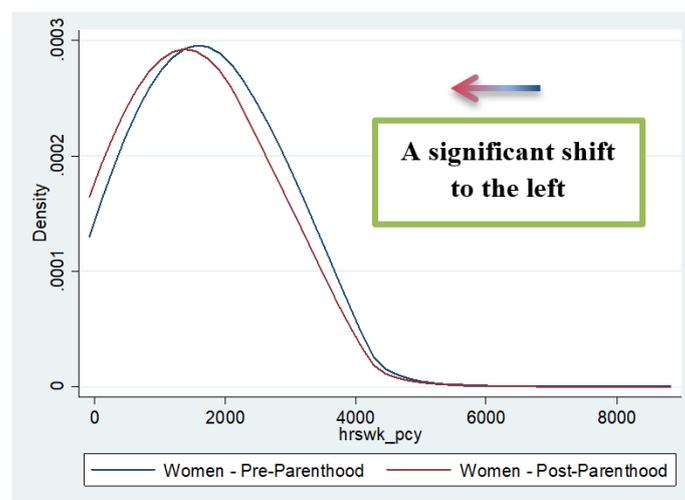

**Figure 2.b & 2.c.** *The Kernel Densities of the Number of Hours Worked for Employed Men (diagram 2.b placed on the left) and Women (diagram 2.c placed on the right) before Becoming Parents and after Becoming Parents, Conditional on Being Employed before and after Becoming a Parent*

Additionally, as these two economic theories predict, the mentioned negative relationship should be stronger among mothers with infants and very young children. In order to develop a deeper understanding of the trend of LFP over time, as the individuals in the sample and their children become older, it would be helpful to look at a historical profile of such trends over time, which is a task to be done in table 7.

The differences across individuals and over time in having or not having children as well as being or not being in the labor force provide the variation needed to assess the sensitivity of LFP behavior to parental status. Now, after having a descriptive analysis on the data, in order to deepen our understanding of the causal effect of parenthood on LFP, we will continue with a simple year-by-year difference-in-difference (henceforth, DiD) analysis of the effect of having the first child, as it is an important predictor of lifetime work experience.[10] The panel data described at the beginning of this section, which allows us to keep track of the same individual over time by assigning each individual a unique id number, provides a great opportunity for us to be able to take an identification approach to identify the causal effect of parenthood on LFP.[11] Accordingly, if having the first child discourages LFP, we should see a decrease in the LFP rates of individuals having the first child. Figure 3 illustrates how this identification approach would work to identify the causal effect of interest in the two-year period of 1979-1980.

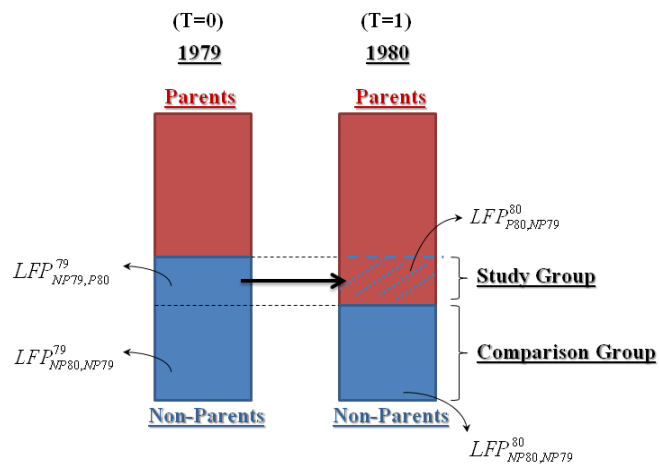

**Figure 3.** *A schematic figure illustrating the study group and comparison group*

---

[10] According to Shapiro & Mott (1994), there is strong evidence that labor force status following the first birth is an important predictor of lifetime work experience.

[11] As mentioned already in the first footnote of the paper, there still exists a subtle difference between the identification approach taken in this paper and the common quasi-experimental identification strategy. This difference mainly has to do with lack of an exogenous source of variation in the identification approach taken in this paper. This shortcoming that has its roots in lack of data on an appropriate instrumental variable may cause spurious effects and some contaminations in the estimates to be made. After all, throughout the paper, it is tired to dispel the mentioned weakness to the extent possible by taking advantage of multiple analytical methods and numerous regression models so that we can gain a better, more comprehensive, and more reliable understanding of what the true causal effect could be.

As illustrated above, naturally, the individuals in the year of 1979 (before treatment) are of two types in terms of their parental status: parents (represented by the red rectangle on the left) and non-parents (represented by the blue rectangle on the left). The same division applies to the individuals in year 1980 (after treatment). However, more interesting to us is a group of individuals who are part of non-parents in 1979, but change their parental status (through the process of childbearing) from non-parents in 1979 to parents in 1980 (these people are represented by two horizontal smaller rectangles, i.e. the blue one on the left that shifts to the hachured, red one on the right). In fact, the LFP behavior expressed by these individuals is of essential importance in identifying the causal effect of parenthood on LFP, as they can be regarded as a treatment group or study group.[12] There is also a group of individuals who are non-parents in 1979 and also remain non-parents in 1980 (represented by the larger red rectangle on the right), which can serve as a control or comparison group in this design. The LFP rates of the study and the comparison groups before and after treatment are denoted as $LFP^{79}_{NP79,P80}$, $LFP^{80}_{P80,NP79}$, $LFP^{79}_{NP80,NP79}$, and $LFP^{80}_{NP80,NP79}$, respectively, which also have been illustrated in figure 3. In such a setting, the causal effect of being a parent (of a first child in the first year) will be equal to the following equation:

$$Causal\_Effect = (LFP^{80}_{P80,NP79} - LFP^{79}_{NP79,P80}) - (LFP^{80}_{NP80,NP79} - LFP^{79}_{NP80,NP79}) \quad (1)$$

Table 6 reports the results of the year-by-year DiD analysis for the two-year period of 1979-1980.

**Table 6.** *Difference-in-difference estimate of the effect of parenthood on LFP*

|  | Average LFP Rate | |
| --- | --- | --- |
|  | Non-Parent Individuals (Treated Group) | Parent Individuals (Control Group) |
| Pre-Treatment (1979) OR (T=0) | 0.6318 (0.0200) | 0.5360 (0.0154) |
| Post-Treatment (1983) OR (T=1) | 0.5988 (0.0220) | 0.5808 (0.0157) |
| Column Difference | - 0.0330 (0.0297) | 0.0448 (0.0220) |
| Difference-in-Difference | - 0.0778 (0.0282) | |

**Note:** Values in parentheses are standard errors of the corresponding point estimates.[13]

---

[12] This identification approach is based on two assumptions. First, there are no other contemporaneous shocks (than the treatment, which is equivalent to childbearing or becoming a parent) to the relative LFP of the study and comparison groups over the course of the study. Second, there are no underlying trends in the labor market and LFP that differ between the study and comparison group.

[13] To estimate the standard error of this DiD estimate, the SE of DiD estimator introduced by Wooldrige (2010, p.148, eq. 6.52) was used. For more information on how the estimates of SEs in this DiD analysis are made, you can see appendix 2.

The results reported in table 6 indicate that the overall average rate of LFP after having the first child will decrease by roughly 8 percentage points. This is indeed the DiD estimate of the effect of parenthood on LFP. This result is both statistically precise and economically significant. After finding such a significant "overall" impact of parenthood on LFP, it makes sense to explore any possible heterogeneous effects of motherhood and fatherhood on the LFP rate. To undertake this task with a historical flavor, equation 1 has been estimated for three different sub-samples called "overall" (aggregated), "male," and "female" in five time periods over the course of the study horizon.[14] The corresponding results are reported in table 7.

**Table 7.** *Year-by-year difference-in-difference estimates of the effect of parenthood on LFP over time*

| | Year | 1979-1980 | 1982-1983 | 1991-1992 | 2000-2002 | 2010-2012 |
|---|---|---|---|---|---|---|
| Effect of Parenthood on LFP (Estimated by Diff-in-Diff Estimator) | *Overall* | -0.078 | -0.071 | -0.022 | -0.006 | -0.014 |
| | | (0.012) | (0.013) | (0.017) | (0.012) | (0.012) |
| | *Women* | -0.141 | -0.133 | -0.014 | -0.052 | -0.040 |
| | | (0.016) | (0.018) | (0.022) | (0.016) | (0.017) |
| | *Men* | 0.082 | -0.001 | -0.001 | 0.025 | 0.02 |
| | | (0.016) | (0.019) | (0.024) | (0.016) | (0.018) |

**Note**: Values in parentheses are standard errors of the corresponding point estimates.

Up to this point, the results listed in tables 5, 6, and 7 imply that there is a highly significant effect of parenthood (especially, of motherhood) on LFP, which can be theoretically justified by having a quick look at the related theoretical literature. As discussed in section 2, the bottom line of both of the traditional neoclassical theory of labor supply as well as Becker's household production model is that motherhood will be negatively related to women's LFP since it increases a woman's reservation wage. In sum, we easily see that the descriptive statistical analysis of the paper (table 5) and the DiD estimates clearly backs up the predictions of the two related economic theories.

Figure 4 depicts the historical trend reported in table 7 in the form of a time profile.

---

[14] The time points of the historical DiD analysis have been selected so for the following reasons. The first and last pairs of years are in fact the two endpoints of the study horizon interval. According to mathews et al (2009), in the 1980s, the average age of a first-time mother was around 22, and this is why the year 1983 has been chosen as one of the years to be studied in the historical DiD analysis. Indeed, this year is selected since the frequency of turning from a non-parent to a parent for the population of the study, whose average age became 22 in 1983, is likely at its maximum, if we assume that it follows a normal distribution. In addition, the other three time periods have been picked so as to compare the LFP rates and trends in almost every ten years. Incidentally, since there were no data for the two years 2001 and 2011, the two years of 2000 and 2010 were selected instead.

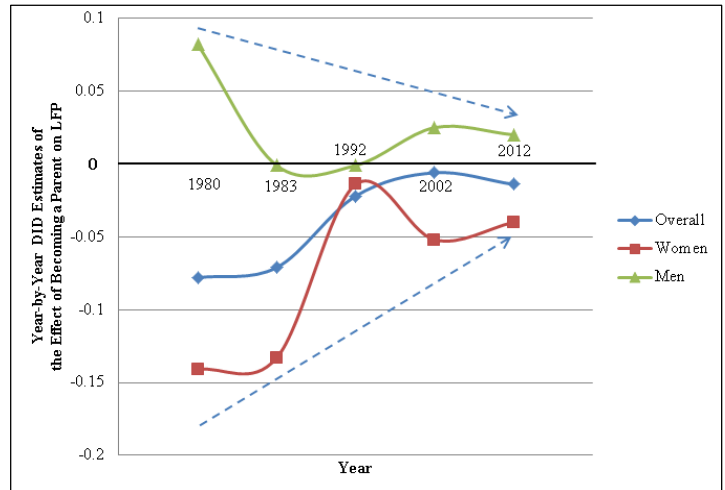

**Figure 4.** *The time profile of the historic trend of the effect of parenthood on LFP. Based on some selected year-by-year difference-in-difference estimates*

In particular, table 7 and figure 4 are very informative in the sense that they reveal two sorts of information. The first is the gender comparison of the effect of parenthood on LFP, and the second is the historical trend of these effects. That is, these show how differentially maternity and paternity affect LFP, and also how they have changed during the past few decades. As the estimated values in the above table and diagram imply, maternity affects LFP negatively and substantially, and the direction of the effect has remained unchanged, while the magnitude of it has dramatically decreased over time. In contrast, paternity has affected LFP often positively and less substantially in the past (relative to that of women), and very trivially in more recent years. Also, the direction of the effect has remained unchanged, while the magnitude of it has dramatically decreased over time. It is also worthwhile to mention that the negative estimates for paternity are not statistically, nor economically significant. These differential effects can simply be justified by the theoretical prediction of Becker's model of household production and work division among family members as discussed in section 2.

Additionally, as these two economic theories predict, the mentioned negative relationship should be stronger among mothers with infants and very young children. In order to develop a deeper understanding of the trend of LFP over time, as the individuals in the sample and their children become older, we looked at a historical profile of these trends over time, which has been reported in diagram 4. As obvious in this diagram, the effect of parenthood on LFP has been fading away over time, regardless of the gender of parents. This phenomenon is completely consistent with the prediction of the mainstream economic theories discussed in section 2. Particularly, as Becker model suggests the negative relationship between motherhood and LFP among mothers with infants and very young children should weaken and finally fade away as the average age of the individuals in the sample increases over time.[15]

The two pieces of evidence mentioned above are empirically amazing and theoretically attractive. This is because these two empirical findings precisely map onto the theoretical predictions that the mainstream economic theory has made about the subject matter at hand.

To conclude the DiD analysis part of the study, it should be noted that the DiD analysis carried out here is of great importance for multiple reasons. First of all, it, by construction, focuses on the short-run effect of having a first child (i.e. the estimated effect of having a child younger than one year old), which is due to the way this DiD analysis was designed. Secondly, as discussed by Cristia (2006) and according to the evidence presented by Shapiro & Mott (1994), an estimate of the impact of having a first child younger than one year old applies to a much wider population than estimates that focus on the effect of a second or higher order child. Thirdly, according to Browning (1992), the short-run impacts of childbearing on female labor supply are substantially larger than the long-run impacts of it. As a logical result of this, knowing the short-run impacts of childbearing is highly informative, since it provides an upper bound for the long-run impacts (Cristia, 2006). By taking into account the three positive points made about the advantages of this DiD analysis, it can be easily understood how important and useful the results of this analysis could be.

Whilst the DiD analysis was straightforward, informative, and useful for the three reasons pointed out above, this simple DiD analysis does not take into account other factors influencing LFP decisions. In particular, other variables – such as age, the number of years of schooling, sex, and marital status – can also have effects on the LFP decision-making process. To account for these factors, univariate and multivariate regression analyses will be carried out in what follows. Also, simple DiD estimates do not exploit all of the available variation in parenthood, which arises both from changes over time and from cross-sectional differences. Hence, it makes sense to use the entire 1979-2012 NLSY79 sample to estimate regression models of LFP, and include multiple control variables to control for different factors potentially influencing LFP decision. Furthermore, to deal with individual-specific and time-specific sources of variations, fixed-effects models along with time dummy variables included in them will be estimated.[16]

---

[15] It is important to note that the fading away of parenthood effect on LFP could also be due to other reasons, which were briefly discussed in sections 2 and 3. As a result, this phenomenon can be empirically attributed to changes in preferences, changes in the costs of childrearing and household maintenance, changes in children's preferences, and the like. However, in the absence of further data, research, and evidence, it is hard to know which one is the main cause of this historical trend. Therefore, since this is beyond the scope of the present paper to further investigate the potential causes of this phenomenon, this topic is left open here for future research.

A wise initial step to start with the regression analysis part of the paper would be constructing the correlation matrix of the data being used. This task has been undertaken in the following table.

**Table 8.** *The correlation matrix of dependent and independent variables*

| Correlations | LFP | Parenthood | HGC | Age | Sex_dummy | Race | Married |
|---|---|---|---|---|---|---|---|
| LFP | 1.000 | | | | | | |
| Parenthood | -0.050 | 1.000 | | | | | |
| HGC | 0.182 | -0.004 | 1.000 | | | | |
| Age | 0.049 | 0.285 | 0.234 | 1.000 | | | |
| Sex_dummy | 0.108 | -0.217 | -0.045 | -0.014 | 1.000 | | |
| Race | 0.065 | -0.061 | 0.158 | -0.059 | 0.000 | 1.000 | |
| Married | 0.049 | 0.497 | 0.127 | 0.249 | -0.056 | 0.092 | 1.000 |

By taking a look at the coefficients of correlation reported above, it becomes apparent that the only pair of explanatory variables that suffers from a fairly considerable collinearity is the pair of the marital status and parenthood with the coefficient of correlation of 0.497. However, since it is not yet considered to be a very strong correlation, these two variables will be kept in some of the regression models to be estimated, although there might exist some confounding effects (because of the multi-collinearity) in the estimated coefficients due to this rather sizable coefficient of correlation.[17] To deal with this potential confounding, three models will be run which exclude the variable "married."[18] Further investigation on the exclusion of the variable "married" did not imply any significant changes in the estimated coefficients. For instance, see regression models 4 and 7, reported in table 9, and compare estimates of these models with those of others.

Based on the nature of the dataset available, the descriptive, statistical investigations, as well as the correlation matrix of the data, fourteen different combinations of regression-model / estimation-methods, including both univariate and multivariate regression models have been estimated through both linear and non-linear estimation methods.[19] The complete results of

---

[16] If one 'truly' sufficiently controls for other influential variables, and also if the key variable of interest possesses an exogenous source of variation, then the "ceteris-paribus effect" estimated can often be considered to be a "causal effect".

[17] In general, in the presence of collinearity, the coefficient estimates of the regression might change dramatically in response to small changes in the data or the model; however, collinearity does not decrease the predictive power and reliability of the estimated model as a whole, and it only might affect individual coefficients estimated, which are individual predictors.

[18] The reason why the variable "married" is tried to be kept in some of the models, while we also have the variable parenthood in the models, lies in the fact that one can simply be single but a parent. There are naturally a number of people who are single parents, so for such realistic cases, the dummy variable married equals 0, while the dummy variable parenthood equals 1.

[19] In terms of estimation methods, this paper employs Binary Response Models (BRMs), including the Linear Probability Model (LPM) as well as two non-linear BRMs, i.e. Logit and

these regressions have been reported in appendix 3 and table 12.[20] Table 9 summarizes these results for a select set of these models.

**Table 9.** *Regression results for a select set of models estimated using various estimation methods*

| Regression Method and Number | Parenthood | Age | Sex_dummy | HGC | Married | R^2 | Description |
|---|---|---|---|---|---|---|---|
| 1. Pooled OLS (LPM) | -0.076 | | | | | 0.02 | Dummies for years included |
|  | (0.002) | | | | | | |
| 2. Fixed-Effects Regression | -0.044 | | | | | 0.11 | Dummies for years included, and individual-specific dummies included |
|  | (0.002) | | | | | | |
| 3. Pooled OLS | -0.062 | -0.0019 | 0.079 | 0.027 | 0.043 | 0.06 | Dummies for years included |
|  | (0.002) | (0.0004) | (0.002) | (0.000) | (0.002) | | |
| 4. Fixed-Effects (Within) Regression | -0.044 | -0.0007 | | 0.021 | | 0.06 | Dummies for years included, individual-specific dummies included, and "married" excluded |
|  | (0.002) | (0.0004) | | (0.001) | | | |
| 5. Probit | -0.235 | -0.009 | 0.310 | 0.113 | 0.172 | 0.06 | Dummies for years included |
|  | (0.008) | (0.001) | (0.007) | (0.001) | (0.008) | | |
| 6. Logit | -0.424 | -0.015 | 0.560 | 0.203 | 0.305 | 0.06 | Dummies for years included |
|  | (0.014) | (0.003) | (0.012) | (0.003) | (0.013) | | |
| 7. Logit | -0.291 | -0.009 | 0.577 | 0.207 | | 0.06 | Dummies for years included, and "married" excluded |
|  | (0.013) | (0.003) | (0.012) | (0.003) | | | |

**Notes**: The dependent variable is LFP. The time unit of observations is annual. All of the above estimated coefficients are statistically significant at p>0.001. The values reported in R^2 column for non-linear estimation models and the corresponding Pseudo R^2's.

The estimates made through both linear and non-linear estimation methods in the above table essentially tell a similar story, suggesting that, for instance, the key independent variable, i.e. parenthood, has had a negative effect on the LFP decisions. Indeed, all the estimates reported in table 9 can be interpreted in terms of probabilistic statements.[21] As you can

---

Probit models, to estimate Labor Force Participation (LFP) equations using individual-level data.

[20] A random-effects model was also estimated, but the results of the related Hausman test and also Breusch–Pagan test were in favor of the compared fixed-effects model.

[21] In fact, the upper part of table 9 reports the results associated with Linear Probability Model (LPM) which essentially uses the OLS estimation method without assuming any distribution over the error term, while the results reported in the lower part of table 9 are

see in table 9, there is a line in the middle of the table that separates linear and non-linear regression estimation methods, indicating that the estimates of those models should not be directly compared with those of the linear ones. The coefficient estimates produced through linear estimation methods have rather clear interpretations. For instance, the estimated coefficient of parenthood in the first model implies that being a parent decreases the likelihood of being in the labor force by 7.6 percentage points (which is a very close estimate to that of DiD analysis for the first two-year period).[22] However, interpreting the estimated coefficients of the non-linear models is essentially more complicated and subtler. In order to be able to interpret these coefficients, one needs to rely on calculus, and first transform the coefficients, and then interpret them. For instance, the estimated coefficient of parenthood in the last logit model[23] (model 7), which equals -0.291, can be interpreted as follows: parenthood has a negative effect on LFP by roughly 4 percentage points.[24, 25] This value is the partial effect at the average (PEA), while the average partial effect (APE) is more comparable to the results from LPM estimates.

Next, on the basis of the results from the various, numerous regression models run for the aggregated data set, a select set of regression models are chosen to be run for the disaggregated data, so that we can somehow distinguish and thereby better identify the differential effects of parenthood on the LFP status for individuals of different gender. The results of these regression models are presented in table 10.

---

associated with the non-linear models, which essentially use the MLE estimation method and implicitly assumes a normal distribution (for probit model) or a logistic distribution (for logit model) over the error term when estimating the coefficients of interest and the causal effect of the key explanatory variable.

[22] An important point to mention here is that these regression models account for being a parent, regardless of the number of children, as long as the children is resident with the parent, and also is younger than 18 years old. This definition is different from the definition used in the DiD analysis, which accounted only for the first child. In some sense, the word parenthood is being defined differently than in the DiD analysis part. However, as will be seen, it is interesting that the results are fairly robust to even the substantial changes in the definition of the word parenthood.

[23] It is important to recognize that this logit model is essentially a "multinomial" logit model, whose explanatory variables are characteristics of the units (here, individuals), not those of the choice to be made. Alternatively, one can work with a "conditional" logit model, in which data consist of choice-specific attributes, such as wage under the LFP setting. As another option, one can choose to run a "mixed" logit model that is a combination of both. After all, doing these is beyond the scope of the present paper, and they are all left for future research.

[24] This value is the partial effect at the average (PEA), i.e. the partial effect of parenthood for the average person in the sample. Alternatively, one can compute the average partial effect (APE), by averaging the individual partial effects across the sample.

[25] After all, the estimates from the three models (LPM, logit, and Probit) tell us a consistent story. As reported in the above table, the signs of the coefficients remain the same across the models, and the same variables are statistically significant in each model. However, as mentioned earlier, the magnitudes of the coefficient estimates cannot directly be compared across models, and one needs to compute the corresponding scale factors so as to be able to compare the coefficient estimates of interest.

**Table 10.** *Regression results for the disaggregated data to study the differential effects of parenthood on LFP for individuals of different gender*

| Regression Method and Number | Variables | | | | | | | | Description |
|---|---|---|---|---|---|---|---|---|---|
| | Parenthood | | Age | | Years of Schooling | | Marital Status | | |
| | Male | Female | Male | Female | Male | Female | Male | Female | |
| 1. Pooled OLS (LPM) | 0.054 (0.002) | -0.158 (0.003) | | | | | | | Dummies for years included |
| 2. Fixed-Effects Regression | 0.032 (0.003) | -0.062 (0.003) | | | | | | | Dummies for years included, and individual-specific dummies included |
| 3. Pooled OLS | 0.027 (0.003) | -0.124 (0.003) | -0.001 (0.000) | -0.002 (0.001) | 0.015 (0.000) | 0.037 (0.001) | 0.043 (0.003) | 0.014 (0.003) | Dummies for years included |
| 4. Fixed-Effects Regression | 0.025 (0.003) | -0.108 (0.003) | 0.000 (0.000) | 0.002 (0.000) | 0.018 (0.001) | 0.024 (0.001) | 0.011 (0.003) | -0.039 (0.003) | Dummies for years included, and individual-specific dummies included |
| 5. Probit | 0.135 (0.015) | -0.466 (0.010) | -0.008 (0.002) | -0.006 (0.002) | 0.075 (0.002) | 0.138 (0.002) | 0.215 (0.014) | 0.045 (0.009) | Dummies for years included |
| 6. Logit | 0.270 (0.028) | -0.813 (0.018) | -0.017 (0.004) | -0.008 (0.003) | 0.138 (0.004) | 0.244 (0.004) | 0.415 (0.027) | 0.090 (0.016) | Dummies for years included |

**Notes**: The dependent variable is LFP. The time unit of observations is annual. All of the above estimated coefficients are statistically significant at p>0.001, except for the estimated coefficients of age in model 4 which are not statistically significant at p>0.001, although they not economically significant, either. To see the numerical values of corresponding $R^2$ and Pseudo $R^2$, see appendix 3 and table 13.

As obvious in the above table, in all of the estimated models, the estimated coefficients of maternity and paternity take on opposite signs, implying differential effects of motherhood and fatherhood on LFP, which are also predicted by the mainstream economic theories. However, the estimated intensities of these differential effects vary across models. For instance, among the linear models, this difference is the largest in model 1 (being 0.212)[26], and the smallest in model 2 (being 0.094).

The estimated coefficients of age are mostly economically insignificant (most of them are very close to zero), suggesting that age does not have any significant effect on LFP in this sample. Although it may seem to be shocking at first glance, it should be recognized that this estimated effect is an intuitive one if one takes into consideration the age structure and the age range of the population under study. That is to say, all of the individuals in this sample have ages ranging from 14 to 55, which almost all considered to be working ages. As a result of this truncation in age, there is no observation corresponding to older people who start to retire and leave the labor force, as already discussed in the descriptive statistical part of the paper.

---

[26] Although you may tend to think these differential effects are not economically so significant, but they actually are. By taking a closer look at the issue, you will see that if you take the difference 0.054 – (-0.158) you will end up with 0.212, and this result and such a large difference economically is a highly significant result, and as such, very interesting in understanding how differentially individuals of different genders arrive at their LFP decisions.

Econometrically, this uniformity does not leave any variation for the variable "age" to show an effect on LFP in this sample.

Arguably, the results for marital status and educational attainment indicate that these two have positive effects on LFP in almost every model. An additional interesting result is that the effect of being married on LFP for men is almost four to five times as large as that for women, depending on the estimated model, which is completely in line with the predictions of the two economic theories discussed in section 2. Moreover, the effect of schooling on LFP for women is almost twice as large as that for men, depending on the model. This is completely intuitive, since having a college degree plays a more important role for women, in making their decision whether or not to participate in the labor force, than it does for men, who usually participate mostly regardless of whether or they have more years of schooling. This empirical result matches up the predictions of Becker's household production model.

To check the validity of the results, a few robustness checks have been conducted. The estimated coefficients reported in table 9 and 10 are all acceptably robust to the application of various estimation methods, different regression models, different functional forms, and inclusion and exclusion of different variables. The main results from these robustness checks and multiple sensitivity analyses, which have been performed to check the sensitivity of the results, show that the results reported in table 9 and 10 are robust to a great extent.

The attractiveness of the results of this study lies in the fact that by focusing on a gender comparison, we can see the causal effect of paternity and maternity, separately, on LFP. In fact, though in many models we saw that the overall effect of parenthood (or becoming a parent) on LFP is overall negative in all the models, when the data were disaggregated by gender, we realized that the overall effect was not saying much about the heterogeneous effects of maternity and paternity. That is, neither does that estimate represent the direction of the effect of paternity on LFP, nor it represents well the magnitude of the effects of maternity on LFP. The same argument applies to the identification of the effects of schooling and marital status on LFP.

In the next section, threats to identification as well as some solutions to overcome these threats are discussed with a reference to the existing literature.

## 5. Threats to identification and suggested solutions

A potential weakness of this paper is that this study does not account for possible reverse causality and endogeneity in the parenthood status. The reason for this is that there was no information on an appropriate IV variable in the data set for it to be used, as far as the author is aware. Despite this, this threat and some potential solutions to deal with this threat are to be explained here, asking for future research to allow for this type of endogeneity in their models.

One may face a challenge in identifying the causal effect of parenthood on LFP, as parenthood may be associated with unobservable parental characteristics and other factors influencing LFP decision, even after controlling for some other observable variables. In other words, although fixed-effects dummy variables, time dummy variables, and controls for other variables can alleviate the omitted variable bias to some degree, potential biases will still remain from reverse causality (where childbearing decisions are made in response to current or anticipated changes in LFP) and joint determination (where childbearing decisions and LFP decisions are made simultaneously).

As Browning (1992) notes in his literature review on the topic, since the childbearing decision may be endogenous, the strong negative correlations found between different measures of fertility and female LFP should not be interpreted as pure evidence of causal effects. The main approach suggested in the literature to account for this endogeneity is instrumental variable (IV) estimation. This approach involves finding an IV which satisfies two requirements. First, the instrument must be highly correlated with childbearing (aka the "relevance assumption"). Second, it must be uncorrelated with the error term in the LFP equation (aka the "validity or exogeneity assumption") (Wooldridge, 2010). According to Van der Stoep (2008), common IVs used to identify an exogenous effect of parenthood on LFP in the related literature are as follows: twin births (Connelly et al, 2006; Bronars & Grogger, 1994), same sex sibling composition (Iacovou, 2001; Angrist & Evans, 1998; Cruces & Galiani 2007; and Van der Stoep, 2008), fecundity or infertility status (Aguero and Marks, 2008) and the Chinese Lunar calendar (Vere, 2008).

As Miller (2009) elaborates, some research studies use socioeconomic background and beliefs as IVs, but deemphasize their IV estimates as imprecise (Blackburn et al., 1993) or unstable (Chandler et al., 1994). However, as she mentions, the potential endogeneity of the instruments and their direct effects on career path and LFP could be even more problematic to the causal identification, than the endogeneity itself would. Hence, one last caveat here is that we, as economists, must eventually make a choice between resorting to a potential endogeneity versus employing a misleading IV variable.[27] In such situations, common sense dictates to adhere to the simplicity of making an exogeneity assumption, and accept the potential for some endogeneity threat, rather than sticking to the fanciness of using IV variables and consequently threatening the internal and external validity of the results.

Finally, it is worthwhile to point out that some scholars, e.g. Cristia (2006) and Van der Stoep (2008), have shown that even when one employs an empirical strategy that accounts for the endogeneity of parenthood, as Cristia (2006) mentions, "interestingly, evidence strongly suggests that the estimates obtained using this strategy (which tackles the problem of the endogeneity) are similar to estimates derived from approaches that assume the exogeneity of childbearing." Also, as Van der Stoep (2008) concluded from her literature review, many studies show that the negative relationship is robust to whether motherhood is treated as exogenous or endogenous with respect to LFP. Such pieces of evidence strengthen the validity and reliability of the results of the present paper. It is also important to remind that no single approach is perfect and ideal, so it is essential for economists to accumulate evidence using various empirical strategies, and this study is just such an attempt to contribute to the related existing literature.

---

[27] Considering the fact that proper IV variables are essentially rare, the IV identification strategies that employ inappropriate IVs can be simply turned into a "stumbling block" rather than a "stepping stone."

## 6. Conclusion

Identifying the factors that influence labor force participation could elucidate how individuals arrive at their labor supply decisions, whose understanding is, in turn, of crucial importance in analyzing how the supply side of the labor market functions.[28] Labor supply decisions, in essence, can be roughly classified under two categories: (1) whether to work at all or not, and (2) how many hours to work, conditional on working positive hours. The present paper is essentially concerned with the former type of decisions.

This paper attempts to isolate the effects of parenthood status on labor force participation (LFP) decisions by exploiting a difference-in-difference identification strategy as well as an individual-level fixed-effects identification strategy. The differences across individuals and over time in having or not having children as well as being or not being in the labor force provide the variation needed to assess the sensitivity of individuals' LFP behavior to parenthood. At first glance, it seems that a change in parenthood status could have totally different impacts on mothers than it would on fathers, for different reasons discussed throughout the paper. Hence, the data is disaggregated by gender in order to be able to look at the effect of maternity and paternity, separately, on LFP.

The primary data source used is the National Longitudinal Surveys (NLS). Considering the nature of LFP variable, this paper employs Binary Response Models (BRMs) to estimate Labor Force Participation (LFP) equations using individual-level (micro) data. The findings of the study show that parenthood has a negative overall effect on LFP. However, paternity has a positive effect on the likelihood of being in the labor force, whilst maternity has a negative impact on LFP. In addition, the results imply that the effect of parenthood on LFP has been fading away over time for the sample of specific individuals under study, regardless of the gender of parents. These two pieces of evidence are empirically amazing and theoretically attractive, since these two empirical results precisely map onto the theoretical predictions made by the related mainstream economic theories (the traditional neoclassical theory of labor supply as well as Becker's household production model). Indeed, the evidence totally backs up the theory. These results are highly robust across different models specified and various estimation methods employed. These findings can contribute to the existing knowledge about the effect of parenthood on LFP decisions made in the US at an individual and behavioral level, and also aid in the shaping of economic policies and interventions to enhance the status of labor force participation in the economy.

---

[28] Moosavian (2016) visually show how labor supply and labor demand are derived, situated, and play roles in the determination of the macroeconomic equilibrium in the economy in a graphical way.

This paper does not account for potential endogeneity of parenthood. The reason for this is that there was no information on an appropriate IV variable in the data set for it to be used, as far as the author is aware. Despite this, this threat and some potential solutions to deal with this threat are explained. However, one last caveat is that we, as economists, must eventually make a choice between resorting to a potential endogeneity versus employing a misleading IV variable. In such situations, common sense dictates to adhere to the simplicity of making an exogeneity assumption, and accept the potential for some endogeneity threat, rather than sticking to the fanciness of using IV variables and consequently threatening the internal and external validity of the results.

Finally, it is worthwhile to point out that some scholars, e.g. Cristia (2006) and Van der Stoep (2008), have shown that even when one employs an empirical strategy that accounts for the endogeneity of parenthood, as Cristia (2006) mentions, "interestingly, evidence strongly suggests that the estimates obtained using this strategy (which tackles the problem of the endogeneity) are similar to estimates derived from approaches that assume the exogeneity of childbearing." Also, as Van der Stoep (2008) concluded from her literature review, many studies show that the negative relationship is robust to whether motherhood is treated as exogenous or endogenous with respect to LFP. Such pieces of evidence strengthen the validity and reliability of the results of the present paper. It is also important to remind that no single approach is perfect and ideal, so it is essential for economists to accumulate evidence using various empirical strategies, and this study is just such an attempt to contribute to the related existing literature.

**Appendices**

**Appendix 1.**

**Table 11.** *Labor Force Participation of the US in years 1994, 2004, 2014*

| Group | Participation rate | | |
|---|---|---|---|
| | 1994 | 2004 | 2014 |
| Total, 16 years and older | 66.6 | 66 | 62.9 |
| 16 to 24 | 66.4 | 61.1 | 55 |
| 16 to 19 | 52.7 | 43.9 | 34 |
| 20 to 24 | 77 | 75 | 70.8 |
| 25 to 54 | 83.4 | 82.8 | 80.9 |
| 25 to 34 | 83.2 | 82.7 | 81.2 |
| 35 to 44 | 84.8 | 83.6 | 82.2 |
| 45 to 54 | 81.7 | 81.8 | 79.6 |
| 55 and older | 30.1 | 36.2 | 40 |
| 55 to 64 | 56.8 | 62.3 | 64.1 |
| 55 to 59 | 67.7 | 71.1 | 71.4 |
| 60 to 64 | 44.9 | 50.9 | 55.8 |
| 60 to 61 | 54.5 | 59.2 | 63.4 |
| 62 to 64 | 38.7 | 44.4 | 50.2 |
| 65 and older | 12.4 | 14.4 | 18.6 |
| 65 to 74 | 17.2 | 21.9 | 26.2 |
| 65 to 69 | 21.9 | 27.7 | 31.6 |
| 70 to 74 | 11.8 | 15.3 | 18.9 |
| 75 to 79 | 6.6 | 8.8 | 11.3 |
| 75 and older | 5.4 | 6.1 | 8 |

**Appendix 2.** *Estimating the SEs of the DiD analysis*

The standard error of the mean (SEM) is the standard deviation of the sample-mean's estimate of a population mean. SEM is usually estimated by the sample estimate of the sample standard deviation divided by the square root of the sample size, and the related formula is as the following:

$$SE_{\bar{X}} = \frac{s_X}{\sqrt{n}}$$

where $s_X$ is the standard deviation of the sample and n is the sample size.

The first difference between the means of two samples, $X_1$ and $X_2$, both randomly drawn from the same normally distributed source population, belongs to a normally distributed sampling distribution whose overall mean is equal to zero and whose standard error is equal to:

$$SE_{\bar{X}_2 - \bar{X}_1} = \sqrt{\frac{s_{X_1}^2}{n_1} + \frac{s_{X_2}^2}{n_2}}$$

where $s_X$'s are the standard deviations of the samples (or more accurately, the SDs of the underlying populations) and n's are the sample sizes.

In addition to the above-mentioned formulas, which essentially produce the estimates of SEs of the point estimations of means and first differences (FDs) of the means of interest, one can run the following regression suggested by Wooldridge (2010, p.148, eq.6.52), and all of them will produce "exactly" the same estimates as obtained from the point estimations method.[29] The only element of table that is somewhat cumbersome to estimate using the above-mentioned simple formulas is the SE of DiD estimate, which can be obtained by looking at the estimated SE of the estimated coefficient $\hat{\delta}_1$ (aka DiD estimator) in the following regression which has been suggested by Wooldridge (2010, p.148, eq.6.52).

$$y = \beta_0 + \beta_1 dB + \delta_0 d2 + \delta_1 d2.dB + u$$

For further information on the above equation and its variables and coefficients, please refer to Wooldridge (2010, p.147-148).

---

[29] Wooldridge (2010, p. 148) explains well why these two methods are essentially the same method.

**Appendix 3.**

**Table 12.** *Details of Regression Results*

| Regression number | Regression Method | Variables | | | | | | | $R^2$ | Description |
|---|---|---|---|---|---|---|---|---|---|---|
| | | Parenthood | Age | Sex dummy | HGC | Married | Constant | | | |
| 1 | Pooled OLS (LPM) | -0.076 (0.002) | | | | | 0.834 (0.001) | | 0.02 | Dummies for years included |
| 2 | Fixed-Effects Regression | -0.044 (0.002) | 0.0004 (0.0000) | | | | 0.713 (0.003) | | 0.11 | Dummies for years included, and individual-specific dummies included |
| 3 | Pooled OLS | -0.046 (0.002) | -0.0019 (0.0004) | 0.083 (0.002) | 0.029 (0.000) | 0.046 (0.002) | 0.399 (0.005) | | 0.05 | No dummies for years included |
| 4 | Pooled OLS | -0.062 (0.002) | 0.0008 (0.0002) | 0.079 (0.002) | 0.027 (0.000) | 0.043 (0.002) | 0.382 (0.009) | | 0.06 | Dummies for years included |
| 5 | Fixed-Effects (Within) Regression | -0.042 (0.002) | -0.0007 (0.0004) | | 0.021 (0.001) | -0.007 (0.002) | 0.470 (0.010) | | 0.02 | Dummies for years included, and individual-specific dummies included |
| 6 | Fixed-Effects (Within) Regression | -.044 (0.002) | -0.0068 (0.0009) | | 0.021 (0.001) | | 0.473 (0.010) | | 0.06 | Dummies for years included, individual-specific dummies included, and "married" excluded |
| 7 | Random-Effects Regression | -0.045 (0.002) | 0.0023 (0.0010) | 0.076 (0.004) | 0.024 (0.001) | -0.001 (0.002) | 0.397 (0.008) | | 0.02 | Dummies for years included |
| 8 | Fixed-Effects (Between) Regression | -0.103 (0.008) | | 0.082 (0.004) | 0.027 (0.001) | 0.096 (0.008) | 0.691 (0.032) | | 0.25 | Dummies for years included |

Note: Values in parentheses are standard errors of the corresponding point estimates.

**Table 12.** *Details of Regression Results (Cont.)*

| Regression number | Regression Method | Variables | | | | | | | Pseudo R^2 | Description |
|---|---|---|---|---|---|---|---|---|---|---|
| | | Parenthood | Age | Sex_dummy | HGC | Married | Constant | | | |
| 9 | Probit | -0.158 (0.008) | 0.002 (0.000) | 0.326 (0.007) | 0.119 (0.001) | 0.179 (0.007) | -0.762 (0.020) | | 0.05 | No dummies for year included |
| 10 | Probit | -0.235 (0.008) | -0.009 (0.001) | 0.310 (0.007) | 0.113 (0.001) | 0.172 (0.008) | -0.744 (0.035) | | 0.06 | Dummies for years included |
| 11 | Probit | -0.158 (0.007) | -0.005 (0.001) | 0.318 (0.007) | 0.116 (0.003) | | -0.836 (0.034) | | 0.06 | Dummies for years included, and "married" excluded |
| 12 | Logit | -0.285 (0.014) | 0.003 (0.001) | 0.590 (0.012) | 0.212 (0.003) | 0.320 (0.013) | -1.475 (0.035) | | 0.06 | No dummies for years included |
| 13 | Logit | -0.424 (0.014) | -0.015 (0.003) | 0.560 (0.012) | 0.203 (0.003) | 0.305 (0.013) | -1.423 (0.061) | | 0.06 | Dummies for years included |
| 14 | Logit | -0.291 (0.013) | -0.009 (0.003) | 0.577 (0.012) | 0.207 (0.003) | | -1.579 (0.060) | | 0.06 | Dummies for years included, and "married" excluded |

Note: Values in parentheses are standard errors of the corresponding point estimates.

**Table 13.** *Details of Regression Results*

| Regression Method and Number | Parenthood Male | Parenthood Female | Age Male | Age Female | Yrs of Schooling Male | Yrs of Schooling Female | Marital Status Male | Marital Status Female | Constant Male | Constant Female | Pseudo R^2 Male | Pseudo R^2 Female | Description |
|---|---|---|---|---|---|---|---|---|---|---|---|---|---|
| 1. Pooled OLS (LPM) | 0.054 0.002 | -0.158 0.003 | | | | | | | 0.709 0.005 | 0.683 0.006 | 0.03 | 0.04 | Dummies for years included |
| 2. Fixed-Effects Regression | 0.032 0.003 | -0.062 0.003 | | | | | | | 0.739 0.004 | 0.810 0.002 | 0.02 | 0.03 | Dummies for years included, and individual-specific dummies included |
| 3. Pooled OLS | 0.027 0.003 | -0.124 0.003 | 0.001 0.000 | -0.002 0.001 | 0.015 0.000 | 0.037 0.001 | 0.043 0.003 | 0.014 0.003 | 0.571 0.011 | 0.291 0.013 | 0.08 | 0.04 | Dummies for years included |
| 4. Fixed-Effects Regression | 0.025 0.003 | -0.108 0.003 | 0.000 0.000 | 0.002 0.000 | 0.018 0.001 | 0.024 0.001 | 0.011 0.003 | -0.039 0.003 | 0.549 0.013 | 0.405 0.014 | 0.07 | 0.16 | Dummies for years included, and individual-specific dummies included |
| 5. Probit | 0.135 0.015 | -0.466 0.010 | 0.008 0.002 | -0.006 0.002 | 0.075 0.002 | 0.138 0.002 | 0.215 0.014 | 0.045 0.009 | -0.123 0.051 | -0.928 0.047 | 0.05 | 0.08 | Dummies for years included |
| 6. Logit | 0.270 0.028 | -0.813 0.018 | 0.017 0.004 | -0.008 0.003 | 0.138 0.004 | 0.244 0.004 | 0.415 0.027 | 0.090 0.016 | -0.321 0.093 | -1.761 0.081 | 0.05 | 0.08 | Dummies for years included |

Note: Values listed below estimated coefficients are the standard errors of the corresponding point estimates.

# References

Blau, F.D., & Grossberg, A.J. (1990). Maternal labor supply and children's cognitive development, *NBER Working Paper*, No. w3536. doi. 10.3386/w3536

Boushey, H. (2008). "Opting out?" The effect of children on women's employment in the United States. *Feminist Economics*, 14(1), 1-36. doi. 10.1080/13545700701716672

Browning, M. (1992). Children and household economic behavior. *Journal of Economic Literature*, 30(3), 1434-1475.

Cahuc, P., & Zylberberg, A. (2004). *Labor Economics*. MIT press.

Cramer, D.W., Walker, A.M., & Schiff, I. (1979). Statistical methods in evaluating the outcome of infertility therapy. *Fertility and Sterility*, 32(1), 80-86. doi. 10.1016/S0015-0282(16)44120-8

Cristia, J.P. (2008). The effect of a first child on female labor supply evidence from women seeking fertility services. *Journal of Human Resources*, 43(3), 487-510. doi. 10.3368/jhr.43.3.487

Day, J.C., & Downs, B. (2009). Opting-out: An exploration of labor force participation of new mothers. *Women (ages 28 to 39)*, 1(1.000), 1-000. [Retrieved from].

Dechter, E.K. (2014). Maternity leave, effort allocation, and postmotherhood earnings. *Journal of Human Capital*, 8(2), 97-125. doi. 10.1086/677324

Del Boca, D., & Locatelli, M. (2006). The determinants of motherhood and work status: A survey. IZA Working Paper, No.2414. [Retrieved from].

Eckstein, Z., & Lifshitz, O. (2011). Dynamic female labor supply. *Econometrica*, 79(6), 1675-1726. doi. 10.3982/ECTA8803

Ehrenberg, R.G., & Smith, R.S. (1994). Modem labor economics. *NY: Harper Collins College Publ*.

Goldin, C. (1990). *The Gender Gap: An Economic History of American Women*. New York: Cambridge University Press.

Goldin, C., & Polachek, S. (1987). Residual differences by sex: Perspectives on the gender gap in earnings. *The American Economic Review*, 77(2), 143-151.

Gronau, R. (1988). Sex-related wage differentials and women's interrupted labor careers-The chicken or the egg. *Journal of labor Economics*, 277-301. doi. 10.1086/298184

He, X., & Zhu, R. (2015). Fertility and female labor force participation: Causal evidence from urban China. *The Manchester School*, 84(5), 664-674. doi. 10.1111/manc.12128

Karimi, A. (2014). Impacts of policies, peers and parenthood on labor market outcomes. [Retrieved from].

Miani, C., & Hoorens, S. (2014). Parents at work: men and women participating in the labor force. [Retrieved from].

Miller, A.R. (2011). The effects of motherhood timing on career path. *Journal of Population Economics*, 24(3), 1071-1100. doi. 10.1007/s00148-009-0296-x

Moosavian, S.A.Z.N. (2016). Teaching economics and providing visual "big pictures. *Journal of Economics and Political Economy*, 3(1), 119-133.

Moosavian, S.A.Z.N. (2016). The visual "big picture" of intermediate macroeconomics: A pedagogical tool to teach intermediate macroeconomics. *International Journal of Economics and Finance*, 8(9), 234-248. doi. 10.5539/ijef.v8n9p234

Moosavian, S.A.Z.N. (2016), Using the interactive, graphic syllabus in the teaching of economics, *American Journal of Business Education*,10(2), 45-64. doi. 10.19030/ajbe.v10i2.9914

Neumark, D., & Korenman, S. (1992). Sources of bias in women's wage equations: results using sibling data*, NBER Working Paper*, No.w4019. doi. 10.3386/w4019

Stafford, F.P. (1987). Women's work, sibling competition, and children's school performance. *The American Economic Review*, 77(5), 972-980.

Van der Stoep, G. (2008). Identifying motherhood and its effect on female labor force participation in South Africa: an analysis of survey data. *MCom thesis, University of KwaZulu-Natal, Durban*. [Retrieved from].

Van der Stoep, G. (2008). Identifying motherhood and its effect on female labour force participation in South Africa: an analysis of survey data. *MCom thesis, University of KwaZulu-Natal, Durban*. [Retrieved from].

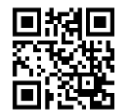